\documentclass[graybox,envcountchap]{svmult}

\newcommand\ho{\ensuremath{H_0}}
\newcommand\mbar{\ensuremath{\overline{m}}}

\newcommand\Mbar{\ensuremath{\overline{M}}}

\newcommand\fbar{\ensuremath{\bar{f}}}

\newcommand{\gsim}{\ \raise-2.truept\hbox{\rlap{\hbox{$\sim$}}\raise 5.truept\hbox{$>$}\ }}
\newcommand{\ksm}{km~s$^{-1}$~Mpc$^{-1}$}
\newcommand\farcs{\mbox{$.\!\!^{\prime\prime}$}}%
\usepackage{mathptmx}        
\usepackage{amsmath}
\usepackage{amssymb}
\usepackage{color}
\usepackage{helvet}          
\usepackage{courier}         
\usepackage{dirtree}

\usepackage{makeidx}        
\usepackage{graphicx}        
\usepackage{subfig}

\usepackage{multicol}        
\usepackage[bottom]{footmisc}

\usepackage{hyperref}        
\hypersetup{colorlinks=true,urlcolor=blue}

\usepackage[misc]{ifsym}

\makeindex             

\begin{document}

\title{Surface Brightness Fluctuations}
\author{Michele Cantiello \& John P. Blakeslee}

\institute{Michele Cantiello (\Letter) \at 
INAF-Astronomical Observatory of Abruzzo, Via Maggini snc, 64020, Teramo, Italy; \email{michele.cantiello@inaf.it}
\and John P. Blakeslee (\Letter) \at NSF's NOIRLab, 
Tucson, AZ 85719, USA; \email{john.blakeslee@noirlab.edu}}
%
%
\maketitle
\vspace{-2.5cm}

\abstract{The Surface Brightness Fluctuation (SBF) method is a powerful tool for determining distances to early-type galaxies. The method measures the intrinsic variance in a galaxy's surface brightness distribution to determine its distance with an accuracy of about 5\%.
Here, we discuss the mathematical formalism behind the SBF technique, its calibration, and the practicalities of how measurements are performed.
We review the various sources of uncertainties that affect the method and discuss how they  can be minimized or controlled through careful observations and data analysis.
The SBF technique has already been successfully applied to a large number of galaxies and used for deriving accurate constraints on the Hubble-Lema\^itre constant $H_0$. 
An approved JWST program will greatly reduce the systematic uncertainties by establishing a firm zero-point calibration using tip of the red giant branch (TRGB) distances.
We summarize the existing results and discuss the excellent potential of the SBF method for improving the current constraints on $H_0$.}

\section{Introduction}

Stellar counts have long been used to study unresolved or partially-resolved stellar populations \cite{baum1955}. However, it was not until the development of CCD astronomy that the statistical characteristics of the discrete nature of stellar counts allowed for the formulation of the surface brightness fluctuations (SBF) method \cite{ts88,tal90}, one of the most reliable and stable extragalactic distance indicators used within the range of a few to about 150 Mpc.

As the distance to a stellar population increases, individual stars blend together and the brightness profile becomes smoother, but there remain statistical fluctuations on small scales because of the discrete nature of the stars.
The SBF method 
leverages the random nature of stellar counts and luminosities to measure a quantity that corresponds to the luminosity-weighted mean brightness of the stars in the stellar population. For evolved stellar systems, this is approximately equal to the mean brightness of the stars on the red giant branch (RGB).

The SBF method was initially developed to determine distances of relatively nearby elliptical galaxies observed from the ground. Over the past few decades, the technique has undergone improvements in the analysis tools and benefited from increasingly advanced astronomical facilities. Consequently, SBF distances  are now derived for a wider class of objects beyond ellipticals: spiral bulges, dwarf galaxies, irregulars, and others. Moreover, the distance limit has increased by a factor of $\sim$7 from the original predictions, and will increase further in the coming years.

Quantitatively, what is measured in the SBF analysis is the ratio between the intrinsic brightness variance (in the absence of blurring) and the average surface brightness over a specific region of a galaxy. This ratio has the units of flux and it is equivalent to the ratio between the second and first moment of the stellar luminosity function within the region analyzed.
Although the definition is simple, there are a number of intricacies in practice.
These involve how to measure the SBF amplitudes in astronomical images, the optimal targets for the measurement, and how to calibrate SBF magnitudes, either empirically or using stellar population synthesis.

In the following sections, we provide a detailed description of the method and summarize the recent results, especially regarding the ``Hubble tension." We conclude with a discussion of the future prospects for the method.

\section{Characterization of the SBF Signal}

The main conceptual difficulty related to SBF does not pertain to the measurement of fluctuations or their calibration, which we discuss below. Rather, a cause of confusion, for some, concerns the nature of the indicator itself. Unlike other commonly used distance indicators like parallaxes, Cepheid variables, and Type Ia supernovae, the SBF signal is not easily ``visualized", and it is not immediately apparent how fluctuations in star counts can be converted into a reliable estimate of distance.

To grasp how the SBF signal is produced and used for distance estimation, we can consider a simplified model of a CCD image of a stellar population unblurred by a point spread function (PSF). Let all the stars have the same (unknown) luminosity $L_*$, with $n_*$ being the average number of stars per pixel and $d$ the distance. Then, $f_*=L_*/(4\pi d^2)$ is the flux contributed per star. We cannot directly measure $f_*$, but we can measure the surface brightness $F_*$ as the mean flux per pixel, $F_*=n_*\times f_*$. Since $n_*\propto d^2$ and $f_*\propto 1/d^2$, $F_*$ is independent of the distance of the stellar system.

We can also measure the standard deviation in the flux per pixel. Because of Poisson fluctuations in the numbers of stars, the standard deviation is $\sigma_{F}=\sqrt{n_*} f_*$.  Now, we define the SBF flux as the ratio of the variance $\sigma_{F}^2$ to the mean:
\begin{equation}
\bar{f}=\sigma^2_{F}/F_* = n_* f_*^2  / (n_*\,f_*)  = f_* = L_*/(4\pi d^2).
\end{equation}
Thus, in this trivial case where the stars are all identical, the SBF flux reduces to the mean flux per star, which depends inversely on the square of the distance.

In reality, stars follow a luminosity function, and a more realistic treatment \cite{moresco22} shows that the ratio of the variance to the mean surface brightness reduces to:
\begin{equation}
\bar{f} = 
\frac{\sum_{i} n_{i} f_{i} ^ 2} {\sum_{i} n_{i} f_{i}} \equiv
\frac{\bar{L}}{4\pi d^2}\,,
\label{sbf.def}
\end{equation}
where $n_i$ is the number of stars of apparent brightness $f_i$,
and $\bar{L}$ is the ratio of the second and first moments of the stellar luminosity function, corresponding to the luminosity-weighted mean stellar luminosity of the population. 
For a more detailed derivation of $\bar{f}$, see \cite{cervino08} or Sec.~3.9.1 of \cite{moresco22}.
$\bar{L}$ is easily calculated from stellar population models; because of the luminosity weighting, it is sensitive to the brightest stars in the population. 
Thus, the SBF flux \fbar, or the equivalent SBF magnitude \mbar, is solely determined by the stellar luminosity function and the galaxy distance.

\section{From SBF Magnitudes to Distances}

Similar to other standard candle methods, deriving SBF distances requires measuring an apparent magnitude, \mbar, and then adopting an absolute \Mbar\ based on a calibration relation to obtain the distance modulus, $\mbar{-}\Mbar$. The following sections describe the process for measuring \mbar\ and estimating \Mbar, as well as the characteristics of suitable target galaxies and the preferred passbands for reliable SBF measurements.

\subsection{SBF measurements step by step} 
\label{sec.measure}

Accurate SBF measurements require correcting and accounting for the different sources of noise or contamination that afflict the signal. In a sense, the fluctuation signal itself is a sort of noise in the star counts per pixel; in the absence of detector noise, photon shot noise, contamination from background sources, and blurring by the PSF, the SBF  would simply be the Poisson variance among pixels due to the varying number and luminosity of stars in each pixel, normalized to the local mean flux of the galaxy.  However, all of these effects are present and must be dealt with.

Since the SBF signal is convolved with the PSF, it can be distinguished from photon and detector noise, which have white-noise power spectra, at least prior to distortion correction.
For this reason, the analysis is done in the Fourier domain, where the SBF signal is measured by determining the amplitude of the variance component in the image power spectrum on the scale of the PSF. Numerous papers have described the steps involved in SBF measurements \cite{tal90,jensen98,blake99,cantiello05,mei05iv,blake10,jensen15,carlsten19,moresco22}. In general, the procedure can be summarized as follows.

After combining individual dithered exposures to make a cleaned, stacked image free from cosmic rays, satellite trails, and detector defects, we estimate and subtract the sky background, mask any bright stars and neighboring galaxies, then derive a smooth isophotal model of the galaxy surface brightness. This model is then subtracted from the image, the remaining point-like and extended sources are masked, and a low-order fit to the background is derived and subtracted to eliminate large-scale residuals from the galaxy model subtraction.
This masking, modeling and subtraction procedure is typically iterative. 
We then detect and mask all compact objects (foreground stars, globular clusters in the galaxy, faint background galaxies) down to a signal-to-noise threshold of $\sim\,$4.5 using an automatic photometry program. These objects, along with any other sources of non-SBF variance (visible dust, brighter satellite galaxies, tidal features, regions of poor model residuals, etc.) are masked after the smooth model and large-scale residuals are removed.

The residual masked frame is then normalized by the square root of the model and contains both the variance from the stellar fluctuations of interest as well as fluctuations from unexcised sources fainter than the detection limit, along with the unconvolved photon and detector noise. The next step is to analyze the power spectrum of the normalized residual masked frame. The fluctuations from the stellar counts and unexcised sources are convolved with the instrumental PSF in the spatial domain; hence, they are multiplied by the Fourier transform of the PSF in the Fourier domain. For accurate SBF measurements, it is crucial to have an accurate characterization of the local PSF. This can be achieved using individual point sources in the field or by utilizing PSF modeling and reconstruction methods.

After azimuthally averaging the power spectrum\footnote{The power spectrum is by definition the squared modulus of the Fourier transform; this explains the normalization of the residual frame by the square root of the galaxy model.}, the amplitude $P_0$ of the astrophysical component of the normalized and masked residual frame's power spectrum $P(k)$ is obtained as the constant term that multiples the ``expectation power spectrum" $E(k)$, which is calculated as: $E(k)=\widetilde{\hbox{PSF}}(k)\otimes \widetilde{\hbox{M}}(k)$, where $\widetilde{\hbox{M}}(k)$ is the power spectrum of the normalized PSF, and $\widetilde{\hbox{PSF}}(k)$ is the power spectrum of the mask window function. There is also the white noise (approximately independent of wavenumber $k$) component $P_1$; thus the power spectrum is modeled as: $P(k)=P_0 \times E(k) + P_1$. 

If the large-scale light distribution is well-subtracted, the power spectrum will be well represented by these two components over most the $k$ range. Additional power may exist at low $k$ (largest spatial scales) due to imperfect galaxy subtraction. The power may also be suppressed at some values of $k$ due to correlation of the pixel noise resulting from geometric distortion correction; this is especially problematic when using linear interpolation of the pixel values. However, for sinc-like interpolation kernels \cite{cantiello05,mei05iv}, the ``damage" to the power spectrum is confined to high $k$ (the smallest scales). In this case, the highest and lowest wavenumbers can simply be omitted from the fit range to ensure an accurate measurement.

It is crucial to note that the fitted $P_0$ value includes contamination from unexcised sources fainter than the detection limit. Any astrophysical source of fluctuation that is not generated from the stellar counts must be estimated and removed by subtracting the contribution from $P_0$. The presence of dust in some galaxies may also affect the power spectrum, and hence the fitted $P_0$, but in ways that depend on the distribution and are not easy to quantify. Optical images and color maps can be used to identify and mask any dusty regions. Typically the SBF measurement is confined to ``clean regions," and we assume all dust patches are identified and masked; thus, no further contribution is considered in $P_0$.

In most cases, the globular clusters (GCs) within the target galaxy are the main source of contamination for $P_0$, with faint background galaxies being a lesser contaminant.
We estimate the extra power from these sources by fitting, down to the detection limit, a combined luminosity function that includes a radially-dependent Gaussian GC luminosity function (GCLF; e.g., \cite{harris01}) and a power-law LF for the galaxies. 
We then estimate the ``residual power" $P_r$ from sources fainter than this limit (which varies with radius) by extrapolating the fitted combined LF, multiplied by the square of the source flux, from the radially-dependent detection limit down to zero flux (see \cite{blake95}). This residual power is then subtracted from $P_0$ to obtain the intrinsic stellar fluctuations, typically denoted $P_f= P_0-P_r$.

\begin{figure}[p]
  \includegraphics[width=0.95\textwidth]{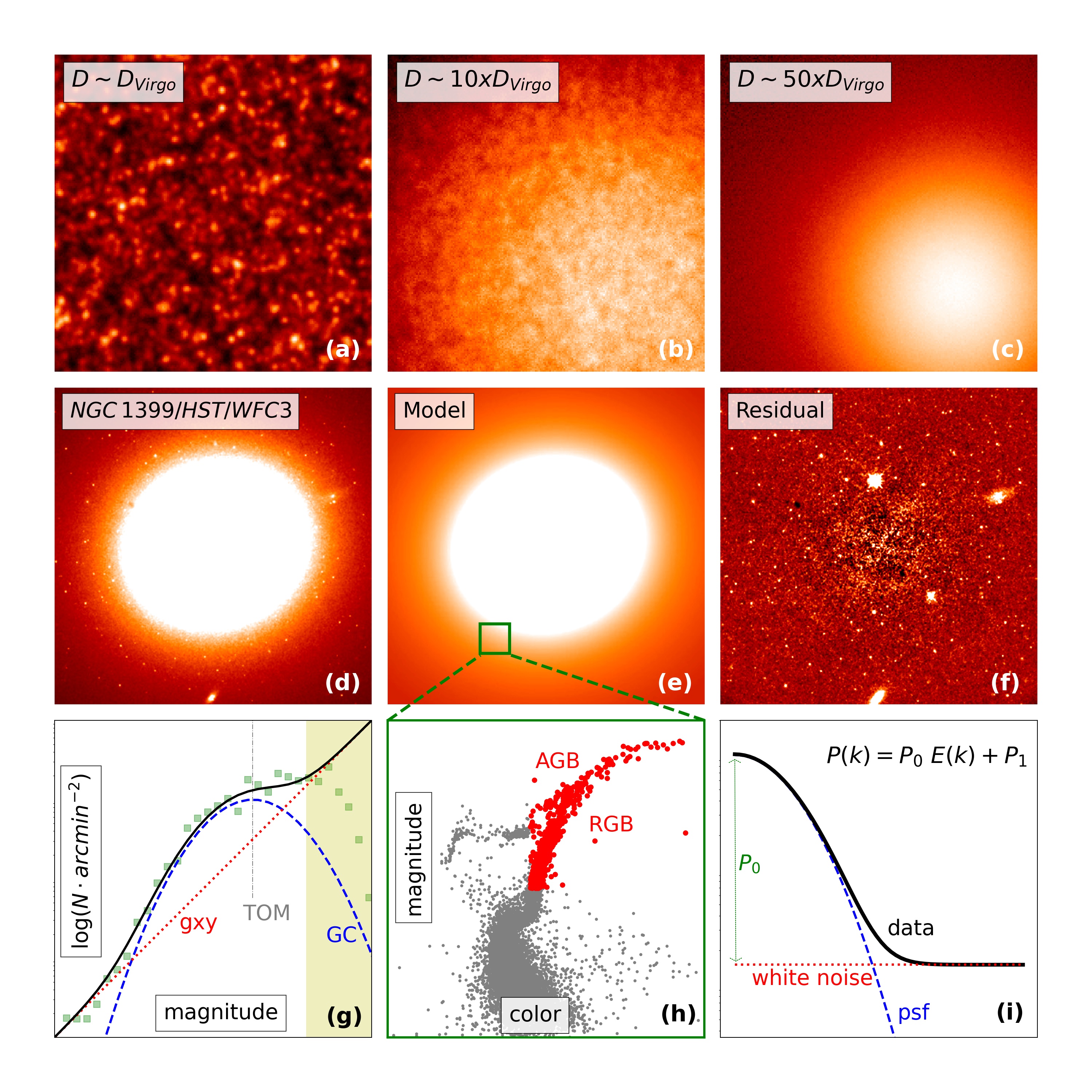}
  \caption{\small Illustration of SBF observations and measurements.  $(a)$~Simulation of the
    stellar population in a spheroidal galaxy at the distance of the Virgo cluster
    ($D_{Virgo}\simeq16.5$ Mpc \cite{blake09}) as observed with the E-ELT in $\sim$1 hour
    (Cantiello et al., 2023, in prep.). $(b)$~Same as in panel $(a)$, but for a galaxy ten
    times more distant. $(c)$~Same as in panel $(a)$, but for a galaxy fifty times more
    distant. Stars, which appear marginally resolved in panel $(a)$, blend together into a
    smooth brightness profile at larger distances. $(d)$~Near-infrared image of NGC\,1399
    from the HST WFC3 camera. $(e)$~Model of NGC\,1399's surface brightness distribution
    derived from the WFC3/IR image.  $(f)$~Residual frame, obtained from the galaxy image
    $(d)$ minus the model $(e)$.  $(g)$~Typical luminosity function analysis for estimating
    the ``residual variance'' $P_r$ due to contaminating sources: green squares show the
    data, the blue curve and red line show the fits to the globular cluster and background
    galaxy luminosity functions, respectively, and the solid black line is the combined
    model luminosity function (data and fits are from\cite{cantiello11}).
    The vertical gray dashed line indicates the GCLF turnover magnitude and the shaded area
    shows the magnitude interval where the detection is incomplete. $(h)$~Color-magnitude
    diagram of an old stellar population (data for the MW globular cluster NGC\,1851 \cite{piotto02}); the RGB/AGB population is highlighted with red dots. $(f)$~A schematic illustration of the SBF power spectrum analysis. Images reproduced with permission from \cite{moresco22,piotto02}.}
  \label{sbf.fig1_sbf}
\end{figure}

Because the GCs have a radial density distribution that increases toward the galaxy center, while the detection limit also gets brighter near the center, the fluctuations from GCs in the central region can overwhelm the signal from the stellar fluctuations, especially for massive galaxies with rich GC systems. Space-based observations (or high-resolution ground-based data obtained with adaptive optics [AO]) can greatly improve the distance limit for SBF by reducing the effect of crowding and pushing the detection limit fainter for the GCs. 
The surface density of background galaxies is typically lower and relatively uniform over the area, which means that their impact is less dramatic and is easier to robustly constrain.

Finally, the SBF magnitude is then calculated as $\mbar =-2.5\log(P_0-P_r)+m_\mathrm{ZP}$, where $m_\mathrm{ZP}$ is the zero point magnitude. For a schematic illustration of the intricate SBF analysis procedure, see Fig.~\ref{sbf.fig1_sbf}.

\subsection{SBF calibration} 

To convert \mbar\ into a distance, knowledge of the absolute SBF magnitude \Mbar\ of the population  is required. 
The most common method for estimating \Mbar\ is through empirical relationships. Studies of large samples of galaxies at known distances or in compact groups have shown that \Mbar\ can be described as a one-parameter family with integrated color (e.g., $V{-}I$ or $g{-}z$) as a useful parameter. 
The ground-based SBF survey of Tonry and collaborators \cite{tonry97,tonry00,blake99,tonry01} found that a linear relation describes accurately the color dependence of $\mbar_I$ in nearby groups and clusters. They also showed that \Mbar\ in the $I$~band is a universal function of color, with an intrinsic scatter in the range of 0.05~to 0.10~mag. Later studies confirmed a similar intrinsic scatter in the calibration relation for red galaxies in suitable passbands \cite{blake09,jensen15},
although for high-quality SBF measurements over a large color range, the \Mbar-vs-color relation is more accurately described by a higher-order polynomial curve \cite{blake09}.

Since the introduction of SBF, many empirical calibrations of \Mbar\ as a function of various optical and near-IR colors have been presented \cite{bva01,blake10,mei07,jensen15}. However, a major drawback of empirical approaches is the need to establish a new calibration whenever the reference passband for SBF or the reference color of the observing campaign changes. An alternative approach is to use SBF magnitude versus color relations derived from stellar population synthesis models \cite{bva01,cantiello03,worthey93}. The advantage of theoretical calibrations is that \Mbar\ can easily be computed for any passband and reference color. Furthermore, fully theoretical calibrations are independent of other distance indicators, avoiding systematic uncertainties associated with the primary indicator. Consequently, theoretically calibrated SBF can be regarded as a primary (though not geometric) distance indicator \cite{blake12b}.

There is generally good agreement in SBF predictions from independent stellar population synthesis models in red optical bands. Discrepancies arise when comparing SBF magnitudes predicted by different models in bluer passbands or in the near-IR, where the models are not yet as robust \cite{jensen15}. These discrepancies can exceed $\sim$0.2 mag. Such a difference is considerably larger than the empirically derived scatter, rendering stellar population model predictions unreliable as an alternative to empirical $\Mbar$. Therefore, except for a few notable cases \cite{biscardi08}, numerically derived \Mbar\ vs color relations serve mainly as validation tools for empirical calibrations.

Before concluding this subsection, it is important to emphasize that accurate distance estimation using $\Mbar$ calibrations requires precise galaxy colors. Hence, a reliable photometric calibration is crucial to ensure consistency and precision in the distances derived.

\subsection{Reference targets and passbands}
\label{sbf.target}
Owing to the definition of SBF and the measurement procedures outlined above, the ideal target galaxy for fluctuation measurements should have minimal dust, a brightness profile that is sufficiently regular, and a relatively homogeneous population of stars with a constant or at least not highly variable stellar luminosity function across its isophotes. This ensures a relatively constant SBF signal across the regions of the galaxy where the measurement is carried out, reducing the intrinsic variation of \mbar\ and resulting in a more accurate distance estimate with lower errors.

In short, the ideal SBF targets are passively evolving, massive red ellipticals that are free from dust contamination and have a well-mixed population of stars with no recent star formation. Empirical calibrations have shown that such galaxies exhibit a tight correlation between absolute SBF magnitude and optical or near-IR colors, enabling SBF distances with a contribution from intrinsic scatter as low as 0.06~mag in certain passbands\footnote{The final distance errors are larger due to measurement uncertainties, as discussed in Sec.~\ref{sec.errors}.} \cite{blake09, blake10, carlsten19}. Stellar population models also support the narrow scatter of \Mbar\ for colors typical of ``red and dead" galaxies with stellar populations dominated by an old, metal-rich component \cite{bva01,cantiello03}.

While the SBF method was originally developed for bright, morphologically regular early-type galaxies, it has more recently been used for a wider variety of stellar systems, including bulges of spirals and the low mass ultra-diffuse galaxies \cite{blake18,cohen18}. 
In contrast to massive red ellipticals, bluer galaxies with intermediate masses have  properties consistent with younger ages, lower metallicities, or both. Galaxies in this color regime have more uncertainties from the age-metallicity degeneracy\footnote{The age-metallicity degeneracy refers to the effect of stellar populations of a specific age and metallicity displaying similar colors, SBF magnitudes, and certain spectral indices as populations with a younger age and higher metallicity, or vice versa \cite{worthey94}.}, which can result in targets with different stellar populations having the same integrated color. However, having the same integrated colors (which are mostly dominated by main sequence stars) does not always mean the same SBF absolute magnitude. This effect implies a larger scatter in the $\Mbar$-vs-color relations as compared to red massive galaxies, leading to larger scatter associated with the derived distances.

Additionally, galaxies with intermediate colors often host an intermediate-age stellar population (3-7 Gyr), which includes many asymptotic giant branch (AGB) stars. These stars are in a bright and rapidly evolving phase, hence their stochastic appearance can make the SBF signal less stable, especially in near-IR bands \cite{raimondo09}. Although AGBs have little impact on integrated colors they represent an additional source of scatter to the SBF in these galaxies. 
As a result of degeneracy effects and AGB stars, in some bands intermediate-mass targets suffer a larger scatter in SBF amplitude at fixed color, which makes it challenging to measure individual distances with an accuracy similar to the red-massive galaxies.

The accuracy of the SBF technique further drops for bluer, low-mass, low surface brightness  galaxies. This is partly because of increased scatter from stellar population effects, including stochastic variations in the AGB stars. Additionally, measuring SBF in diffuse blue galaxies is challenging because low stellar densities make it harder to achieve an adequate signal-to-noise for the stellar-count fluctuations. As a result, a precision of approximately 15\% should be considered the norm for such galaxies under optimal conditions \cite{greco21}.

Despite these challenges, the development of large-format imagers and wide-field surveys on 4m-class telescopes has increased the interest in SBF as a tool for determining distances to smaller galaxies. Recent studies have applied the SBF technique as a distance indicator for galaxies with surface brightness levels as low as $\mu_r\sim24$ mag/arcsec$^2$ \cite{greco21,carlsten19,kim21,blake18}. In these cases, the technique is typically not exploited for its accuracy but rather because these faint and diffuse galaxies do not offer many reliable alternatives for measuring distances. Despite the larger uncertainty, the SBF technique represents a unique opportunity to confirm galaxy associations and group or cluster membership for candidate satellite galaxies.

Concerning the preferred passbands, SBF distances have been measured in various filters, ranging from visible to near-infrared wavelengths. Shorter wavelength bands are generally avoided because the SBF magnitudes are much fainter and more susceptible to contamination from GCs and dust. Nevertheless, SBF data in bluer bands can be valuable tracers for studying unresolved stellar populations due to their sensitivity to stellar population properties \cite{worthey93,bva01,cantiello03,rodriguez21}. Redder optical ($I$ and $z$) and nearer IR ($YJH$) bands are instead preferred for distance measurements.

Empirical studies have shown that the intrinsic scatter of \Mbar\ at a fixed galaxy color is around $\sim0.06$~mag \cite{blake09,blake10} for the most massive ellipticals in optical bands. In the near-IR bands, the intrinsic dispersion of \Mbar\ is closer to 0.10~mag due to the larger variance arising mainly from bright AGB stars \cite{jensen15,cantiello18}.
Despite the somewhat larger instrinsic scatter, near-IR bands such as $J$ and $H$ have several advantages. The SBF is intrinsically brighter, up to three magnitudes brighter in $K$ than $I$, although the intrinsic scatter in $K$ is significant and not well characterized. Near-IR bands also offer a much more favorable contrast with respect to GCs. Reliable \mbar\ measurements need to mask GCs to at least $\sim0.5$ mag fainter than the GCLF peak in the $I$ band, while contamination can be reduced to the same level in the $K$ band by reaching $\sim\,$2~mag \textit{brighter} than the GCLF peak. Finally, the impact of residual dust contamination in the near-IR is negligible in most cases \cite{luppino93}.

The region around $\sim\,${1}$\,\mu m$ ($Y$ band) is particularly interesting for SBF. According to independent stellar population models, and as expected from the inversion of the slope of the empirical SBF vs color relation between optical and near-IR bands for intermediate-color evolved galaxies, this wavelength should exhibit nearly complete degeneracy with respect to stellar population: \mbar\ remains constant with color, making it an ideal distance indicator \cite{worthey93,cantiello03}. In practice, this does not hold for the reddest, most massive ellipticals, for the which the SBF continues to get fainter with color, but these galaxies are comparatively rare. For future wide-area surveys the $\sim Y$ band may be the best choice for SBF measurements, although further study is needed.

In summary, there not a definitive spectral ``sweet spot" for measuring SBF, and any passband from $\sim\,$0.8~to $\sim\,$2~$\mu m$ has its pros and cons. With the rapid development of near-IR technologies and the significant investment in future astronomical facilities in these passbands (e.g., Roman, the ELTs, and the recently launched JWST), along with the brighter intrinsic luminosity and negligible effect of dust contamination, it seems likely that the future of the SBF method, in terms of distance/depth and number of targets reached, lies primarily in the near-IR.

\section{Uncertainties}
\label{sec.errors}
\subsection{Statistical uncertainties}

Statistical uncertainties affecting SBF can be grouped into three main categories: \mbar\ measurement errors, uncertainty in \Mbar\ due to photometric errors in the galaxy color and intrinsic stellar-population scatter in the calibration relation, and uncertainties from data processing and calibration. The last category includes effects like flat fielding and uncertainty in the photometric calibration. While not unique to SBF, it is important to note their presence since they can impact various parts of the SBF measurement process in non-trivial ways. The uncertainty in the flat-fielding multiplicative factor affects the \mbar\ measurement affecting both the variance and average term in eq. \ref{sbf.def} and the color used in the calibration equation. Any sources of uncertainty in the reduction stage impact in a relatively intertwined way the final uncertainty on the distances from SBF. However, these uncertainties are typically well-controlled with standard data reduction and  calibration procedures, with a typical level of $\sim1\%$. Regarding SBF measurement uncertainties, these are directly tied to the specific steps required for the measurement. Usually, each  error term introduced during any of the various analysis stages described in section \ref{sec.measure}, can be independently estimated and then combined in quadrature with the others. The most relevant error terms are the following.

\begin{itemize}
    \item {\it Sky background.} The process of sky subtraction has an associated error due to spatial and/or temporal variations in the sky emission (especially in the near-IR, where the sky background varies on the time scale of the observations), scattered light, and the extended nature of the galaxy, especially if the field of view is comparable to the galaxy size. The effect of the sky uncertainty increases significantly in the outer parts of the galaxy and for galaxies of lower surface brightness. To determine the final error associated to the sky uncertainty, the entire SBF measurement may be repeated by offsetting the background by the uncertainty level; this alters the final \mbar, and the size of the change is then taken into account for the final uncertainty. The sky uncertainty typically contributes $\sim0.02$ mag of the total error on \mbar\ in the near-IR, and less than that in optical bands;\smallskip
    \item {\it Point Spread Function.}  Accurate characterization of the PSF is essential for a reliable SBF measurement, and this depends on the number of bright, unsaturated stars in the field. Thus, for a given galaxy, the quality of the PSF template depends on the size of the field, stability of the PSF across the field, signal-to-noise, saturation level, etc.
    Once a reliable set of reference stars is identified, the SBF measurement is repeated using each available PSF and possibly a composite PSF model. The standard deviation of these measurements, typically less than $\sim0.03$ mag, is then used as the PSF-fitting uncertainty;\smallskip
    \item {\it Power Spectrum Fitting.}  The accuracy of the power spectrum fit is affected by several factors, such as the number of good pixels in the annulus, the spatial structure of the mask, and the presence of patterns in the residual image caused by bars, shells, tidal interaction features or spiral arms. Low- and high-wavenumber power excesses or deficits can be filtered out during the power spectrum fit, yet the range of useful $k$-numbers needs to be large enough for reliable fitting. The fitting of the PSF power spectrum to the galaxy-subtracted data involves a statistical fit uncertainty of $\sim$0.02 mag.\smallskip
    \item {\it Residual Variance $P_r$.} Various approaches have been used to determine the uncertainty in correcting for variance from GCs and background galaxies that are too faint for direct detection. One approach is to examine the change in $P_r$ for different LF parameters and source detection thresholds. Another involves using different, independent detection and photometry tools (SExtractor, DAOphot, DoPHOT, etc.). Based on a variety of tests, a typical uncertainty of about 25\% is adopted for the $P_r$ estimate; thus, to ensure the contribution to the total error on \mbar\ from the $P_r$ correction is $<0.05$~mag requires that $P_r/P_f\sim0.2$. To keep this error term at an acceptable level, it is essential for the images to have sufficient depth; this is vastly easier to achieve with the superior resolution and reduced background of space-based observations.
    \smallskip
    \item {\it Extinction.} Another component in the error budget is the uncertainty in the extinction correction, typically about 10\% of the extinction correction itself. This is added in quadrature to the other sources of statistical error.
\end{itemize}

As for uncertainty in the adopted value of \Mbar\ for a given galaxy, apart from the systematic zero-point error from the distance calibration discussed in the next section, the statistical error in $\Mbar$ includes terms arising from the error in the integrated color of the galaxy (needed to estimate \Mbar\ from the calibration relation) and from the intrinsic, irreducible scatter in the calibration equation. The latter effect results from stellar population variations and varies with the bandpass.

Overall, SBF measurement errors for well designed observations can achieve \textit{statistical} uncertainties of $\sim\,$0.05 mag in \mbar. For a red galaxy observed in a filter close to 1 $\mu$m, the intrinsic scatter in \Mbar\ is $\sim\,$0.06 mag. Hence, when combined, these errors result in a purely statistical uncertainty that can be as low as 0.08 mag, or 4\% on the distance. Nonetheless, fainter, blue galaxies and sub-optimal observing conditions will have larger statistical uncertainties.

\subsection{Systematic uncertainties}

The main systematic uncertainty in SBF distances is in the \Mbar\ zero point\footnote{As future facilities become capable of reaching distances well beyond 100 Mpc, SBF $k$-corrections that take the galaxy spectrum redshift properly into account may be necessary to avoid introducing further systematic effects \cite{jensen21}. As of now, the 100 Mpc limit is usually not surpassed, so the correction does not pose any problems when addressing systematic uncertainties.}.
Empirical calibrations of \Mbar\ are typically based on the zero point from the SBF ground-based survey by Tonry and colleagues, which obtained SBF for bulges of spiral galaxies with distance estimates from Cepheids \cite{tonry01}. A recent reassessment of this calibration reports a total systematic uncertainty of 0.09 mag in the Cepheid-based SBF distance zero point \cite{blake21}. This includes contributions from the tie between near-IR and optical SBF distances, the tie between SBF and Cepheids, and the current uncertainty in the Cepheid zero point, taking into account the improved distance to the Large Magellanic Cloud from detached eclipsing binaries \cite{pietrzynski19}.

The largest source of systematic uncertainty in this calibration is the tie between SBF and Cepheids. This can be improved to a limited degree with further refinements in Cepheid luminosities with Gaia parallaxes (e.g., \cite{clementini19}) and by using more modern Cepheid distances, including a recently revised distance to M31 (one of the SBF calibrators) with a precision better than 2\% \cite{li21}.
New space-based SBF measurements for all the calibrator galaxies would also help.
However, it is important to note that the ideal galaxies for measuring SBF (red ellipticals, Sec.~\ref{sbf.target}), will never have Cepheid distances because they have no recent star formation. 

An alternative to calibrating SBF from Cepheids is to use the tip of the red giant branch (TRGB) method. The TRGB method is ideal for measuring distances to early-type galaxies and can be calibrated using geometric distances from Gaia (e.g., \cite{dixon23,li23}). Unlike the case for Cepheids, the stellar populations underlying both the SBF and TRGB methods are the same, namely old low-mass stars.  Given this, TRGB distances are the natural choice for calibrating SBF. In an early effort to calibrate SBF using TRGB, \cite{mould09} examined a set of 16 galaxies within 10~Mpc. These were predominantly blue, low-mass early-type galaxies, and thus the scatter was large, although the zero point was consistent with the Cepheid calibration. 

A high-quality TRGB-based SBF calibration must be based on massive red early-type galaxies.
Recently, \cite{blake21} compared SBF and TRGB distances to the few such galaxies for which both types of measurment exist, including the bright ellipticals M\,60 and M\,87 in Virgo, and the dusty merger remnant NGC\,1316 in Fornax. The authors found that the mean offset between the TRGB and Cepheid-calibrated SBF distances of the two galaxies in Virgo was only ${-}0.01$ mag. NGC\,1316, a far less ideal galaxy, had a larger offset of 0.14~mag. Although preliminary, this result shows no significant difference between Cepheid- and TRGB-based SBF calibrations.

With the launch of JWST, the prospects have never been better for improving the SBF zero-point calibration using only modest amounts of observing time. A recently approved Cycle~2 program will measure TRGB distances and SBF magnitudes for 14 luminous, red early-type galaxies reaching out to 20~Mpc. The sample is designed to produce an absolute calibration for the SBF method of better than 2\%. Further reduction in the zero-point uncertainty can be achieved by combining the TRGB- and Cepheid-based calibrations.

In summary, the current systematic uncertainty on SBF distances using the Cepheid-based calibration is 0.09 mag or about 4.2\%. This value could be improved by obtaining new SBF measurements for the bulges of spiral galaxies that host Cepheids, but this approach will never be ideal because the best targets for the SBF method are massive red ellipticals. The preferred route for substantially reducing the systematic uncertainty in \Mbar\ is via the the TRGB method, which relies on the same underlying stellar population as the SBF method. A new JWST program, measuring both the TRGB and SBF  for a well-selected set of early-type calibrators, should soon decrease the systematic uncertainty in SBF distances to below 2\%. Further improvements can be achieve with additional TRGB calibrators and/or combining the TRGB and Cepheid calibrations.

\section{Hubble-Lemaitre Constant}
\subsection{Current constraints from SBF}

Since its introduction for measuring extragalactic distances, the SBF method has been applied to about 400 galaxies. The random uncertainty associated with these distances varies greatly, from $\sim\,$4\% for space-based observations of massive, red galaxies, to $\sim\,$20\% for dwarf galaxies observed from the ground. The distances obtained have been used for a variety of scientific cases. For example, SBF has been used to measure distances to:
\vspace{-1pt}
\begin{itemize}
\item
NGC\,4993, the lenticular galaxy host of the gravitational wave event GW\,170817 which had the first observed electromagnetic counterpart \cite{cantiello18};
\smallskip
\item M\,87, the cD host galaxy of the first supermassive black hole to have its ``shadow" imaged using the EHT \cite{ehtVI};\smallskip
\item hundreds of other member galaxies in the Virgo and Fornax clusters to explore the substructure in these environments \cite{blake09,cantiello18b,dunn06,mei07};\smallskip
\item
numerous ultra-diffuse galaxies \cite{cohen18}, including NGC\,1052-DF2, the galaxy proposed to be devoid of dark matter \cite{blake18,vdk18}.
\end{itemize}
In addition, as the method reaches into the Hubble flow, many authors have used this technique for the estimation of the Hubble constant.

The first study to estimate $H_0$ directly from SBF distances decisively in the Hubble flow used a sample of 15 galaxies from 40 to 130~Mpc observed with NICMOS on HST and obtained $H_0\approx76$ \ksm\  (dropping to $\sim\,$72 when restricted to the six most distant targets).\footnote{These numbers should be updated to the most recent Cepheids calibration, which has a minor impact on the final $H_0$ value.} Another early direct measurement of \ho\ from SBF  used archival HST/ACS observations combined with a theoretical calibration based on population models and likewise found $H_0=76$ \ksm\ \cite{biscardi08}.

Among recent results, we mention three studies that are closely related to each other. 
The first study, by \cite{khetan21} (hereafter K21), presented a re-calibration of the peak magnitude for a sample of 24 local Type Ia supernovae (SNe\,Ia) using SBF distances and a hierarchical Bayesian approach. The authors then extended the calibration to a sample of 96 SNeIa at redshifts $0.02 < z < 0.08$ and obtained $H_0 = 71.2 \pm 2.4 \,(\rm{stat}) \pm 3.4 \,(\rm{sys})$ \ksm, where we have updated the value to use the improved LMC distance from \cite{pietrzynski19} for consistency with the other recent $H_0$ values discussed here.
K21 used a diverse sample of SBF distances, which were collected from observations over a span of twenty years, along with an equally diverse sample of SNeIa. While normalizing the SBF distance sample to a common reference calibration was difficult, the heterogeneity of the sample provided some benefits, as it should in principle be less prone to systematic effects, such as those related to specific instruments or data analysis procedures. However, this also resulted in large final statistical uncertainties.
The value of $H_0$ reported by K21 was roughly midway between those found directly from Cepheid-calibrated SNe\,Ia \cite{riess19} and predicted from  analysis of the cosmic microwave background (CMB) \cite{planck20}. 

In another recent study, Blakeslee et al.\ \cite{blake21} (hereafter B21) used the catalog of SBF distances from Jensen et al.\ \cite{jensen21} to estimate $H_0$. This sample comprises 63 bright early-type galaxies observed with the WFC3/IR instrument on the HST, reaching distances up to 100 Mpc. The resulting value of $H_0 = 73.3 \pm 0.7 \,\mathrm{(stat)} \pm 2.4 \,\mathrm{(sys)}$ \ksm\ agrees well with most other direct measurements in the local universe, but  deviates significantly (by about 2.5\,$\sigma$, including the systematic uncertainty) from the CMB prediction, assuming the standard $\Lambda$CDM model. It is noteworthy that the random uncertainty is about one third of that reported in K21. 

To ensure the reliability of their findings, B21 conducted an analysis of $H_0$ involving four different treatments of galaxy velocities, including group-averaged CMB frame velocities, individual CMB velocities, and flow-corrected velocities from two different models. Additionally, their final $H_0$ combines the mutually consistent SBF calibrations from Cepheids and the TRGB in order to reduce the final systematic error down to the $\sim\,$3.3\% level. As noted previously, the number of TRGB calibrators in this analysis was very small, and future studies incorporating a larger sample of giant ellipticals with TRGB distances measured by JWST and tied to Gaia parallaxes should vastly reduce this systematic error.

The third recent work \cite{garnavich22} related to $H_0$ combines an approach like that of K21 with the WFC3/IR SBF distances from \cite{jensen21}. The authors of \cite{garnavich22} used SBF distances for 25 host galaxies of SNe\,Ia to provide an absolute calibration for the Pantheon+ compilation \cite{scolnic22} of 1500 SNe\,Ia extending far out into the Hubble flow. Using the original Pantheon+ ``SALT2" parameters for the dependence of peak luminosity on decline rate and color, these authors find
$H_0 = 74.6 \pm 0.9 \,\mathrm{(stat)} \pm 2.7 \,\mathrm{(syst)}$ \ksm, which agrees to within ${\sim\,}{1}\sigma$ with the value obtained by B21 (the systematic error is common between the two studies, so the comparison must be done with respect to the statistical error only). 

However, it is important to note that the SNe\,Ia whose luminosities can be calibrated with SBF distances are generally fast decliners in massive host galaxies. When \cite{garnavich22} limit their analysis to fast-declining SNe\,Ia in massive hosts and rederive the SALT2 parameters, the SBF-calibrated $H_0$ from these SNe becomes $73.3 \pm 1.0 \pm 2.7$ \ksm, which is identical to the value derived by B21 from SBF distances in the Hubble flow. This $\sim\,$1$\sigma$ change in $H_0$ suggests a slight difference between the best-fit SALT2 parameters for the relatively small set of fast-declining SNe\,Ia in massive galaxies as compared to the full Pantheon+ sample.

\subsection{Forecasts for new and future facilities}

As outlined in the previous section, recent SBF-based estimates of $H_0$ align more closely with the results found by Cepheid-calibrated SNe\,Ia  than with the predictions from the CMB. The potential use of SBF, particularly with JWST and upcoming astronomical facilities, offers a promising route to derive a more precise and accurate value of $H_0$ that is fully independent of SNe\,Ia and Cepheids.
Here, we present the expected results of the SBF method for estimating $H_0$ using various existing and forthcoming facilities: JWST, Rubin Observatory,  ESA's Euclid mission, the Roman Space Telescope, and the Extremely Large Telescopes (ELTs).

To evaluate the potential for constraining $H_0$ from SBF with each new or forthcoming facility, we proceed as follows. First, we estimate the maximum distance $d_{Max}$ that can be reached with the specific facility, as explained below. 

Then, we count the number of massive elliptical galaxies that can be observed by the telescope/instrument within a distance range of $d_{min}\leq D \leq d_{Max}$. Here, we assume $d_{min}=40$ Mpc to exclude galaxies in the local universe where flow motions and peculiar velocities are large compared to distance errors.
To estimate $d_{Max}$, we assume (as is generally the case) that contamination by GCs is the limiting factor for well-designed SBF measurements.  Thus, $d_{Max}$ is the distance to which the GCs can be detected and removed to a faint enough limit, and the residual contamination reliably estimated, so that the error in the correction drops below the intrinsic scatter in the method. For instance, in the $I$ band, where the peak of the GC luminosity function (GCLF) is at $M_I\approx-8.0$ AB mag, accurate SBF measurements require detecting and removing sources to +0.5~mag fainter than the GCLF peak, or $M_I\approx-7.5$ AB mag. In the $H$-band (which has slightly larger intrinsic scatter), the same relative amount of contamination from GCs can be reached for a detection limit about 1.5~mag \textit{brighter} than the GCLF peak; \cite{jensen15}).
We estimate the near-IR GCLF peak magnitudes from \cite{nantais06} and convert to the AB system.

The expected depth is determined based on available information for the specific facilities. For instance, in the case of Rubin Observatory's LSST, we use the expected 10-year $5\sigma$ point-source depth for the $i$~band, with a GCLF peak of $\sim-8.0$ AB mag and sky coordinates with $\hbox{Dec}(\hbox{\small J2000})\leq 15^{\circ}$.

Another important factor is the selection of an appropriate list of targets. We use the 2MASS Redshift Survey (2MRS), which comprises approximately 43,000 galaxies brighter than $K_s$ = 11.75 mag (Vega) \cite{huchra12}. This survey is nearly complete, with the exception of regions near the Galactic plane, which account for only 9\% of the sky, and an $L^*$ galaxy is included if it is within $\sim$135 Mpc. We first focus on ``optimal" targets, i.e., bright galaxies with a morphological $T$-type\,$\leq-4$ (indicating morphologically regular ellipticals) and an absolute $K$-band Vega magnitude $M_K\leq-25$ mag (similar to M87), where distance is estimated from the 2MRS redshift assuming $H_0=75$~\ksm. 
Under these assumptions, Table~\ref{tab:sbf_h0} (column~7) reports the numbers of optimal targets reachable by each of the new and upcoming facilities mentioned above.

Now, if we relax the selection somewhat on the galaxy morphology and luminosity, accepting $T\leq-1$ and $M_K\leq-22.5$ mag as ``good" candidates, then the numbers of SBF targets increase dramatically, as again shown in column~7 of the table. It is important to note that these numbers represent conservative lower limits. For reference, the NGVS survey on the CFHT, with a median $i$-band seeing of $0\farcs7$ and 5$\sigma$ depth three magnitudes fainter than the GCLF peak, enabled SBF measurements for about 300 galaxies in the Virgo cluster \cite{cantiello18b} down to $M_K\sim -16$ Vega~mag. It was also possible to measure SBF in many galaxies with positive T-type, such as bulges of spirals and galaxies with some dust contamination, thanks to the availability of $u$-band data. Thus, the numbers of candidates measurable with the facilities specified in Table~\ref{tab:sbf_h0} could increase by a factor of a few, or even orders of magnitude, when fainter and more irregular targets are considered. 

For estimating the constraints on $H_0$, we also need to adopt values for the expected distance errors. Thus, for ground-based observations in natural seeing (Rubin/LSST), we conservatively assume a total error in the SBF distance of 8\% for the optimal target galaxies, and 16\% for the ``good" targets. For high-resolution space-based and AO-assisted observations, we assume total errors of 4\% and 8\% for the optimal and ``good" targets, respectively.

All the necessary parameters required for estimating the number of reachable targets with the selected facilities are listed in Table~\ref{tab:sbf_h0}, along with the estimated maximum distances and sample sizes for each facility in selected bands. Taking Rubin as an example, it is expected that approximately 50 optimal targets within $d_{Max}\approx70$ Mpc will be measurable in the Baseline survey area, resulting in a constraint on $H_0$ of $\sim1.2\%$ (stat). For ``good" targets, the estimated number increases to 1500, and the statistical error drops well below the systematic uncertainties from the distance calibration and possible large-scale flow motions.

The Roman Space Telescope is especially promising for future constraints on $H_0$ using SBF because of its combination of HST-like resolution, excellent PSF stability, near-IR coverage, and planned large-area surveys \cite{blake23}. While awaiting the Roman telescope, however, the JWST has unsurpassed capabilities for targeting galaxies one at a time out to $\sim\,$300~Mpc, an unprecedented distance for SBF that exceeds the limit of claimed ``Hubble bubbles" (e.g., \cite{keenan13,romano18}). In the NIRCam passbands reported in Table~\ref{tab:sbf_h0}, we find that around a thousand candidates from the 2MRS catalog will be optimal targets for JWST. Of course, only a few dozen are needed to bring the statistical error comfortably below 1\%, so there will be no shortage of targets from which to choose.

To demonstrate the feasibility of SBF measurements with JWST/NIRCam, we conducted a preliminary analysis on NGC\,7317, the elliptical galaxy in Stephan's Quintet at a distance of about 100~Mpc, using the public Early Release Data.

\begin{figure*}[t]
    \centering
    \includegraphics[width=1.\hsize]{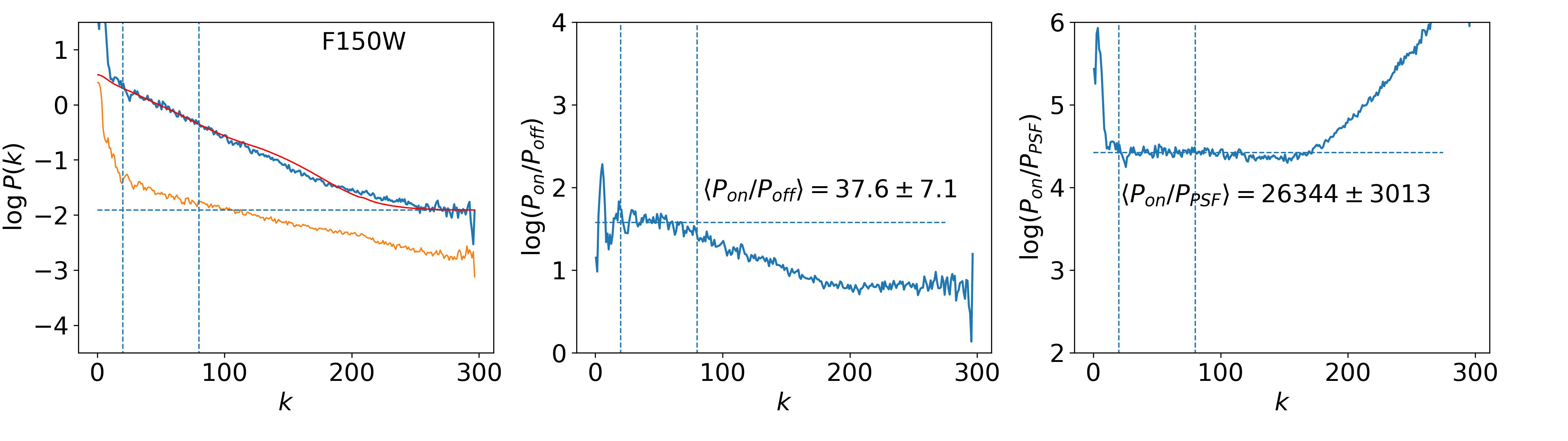}
    \includegraphics[width=1.\hsize]{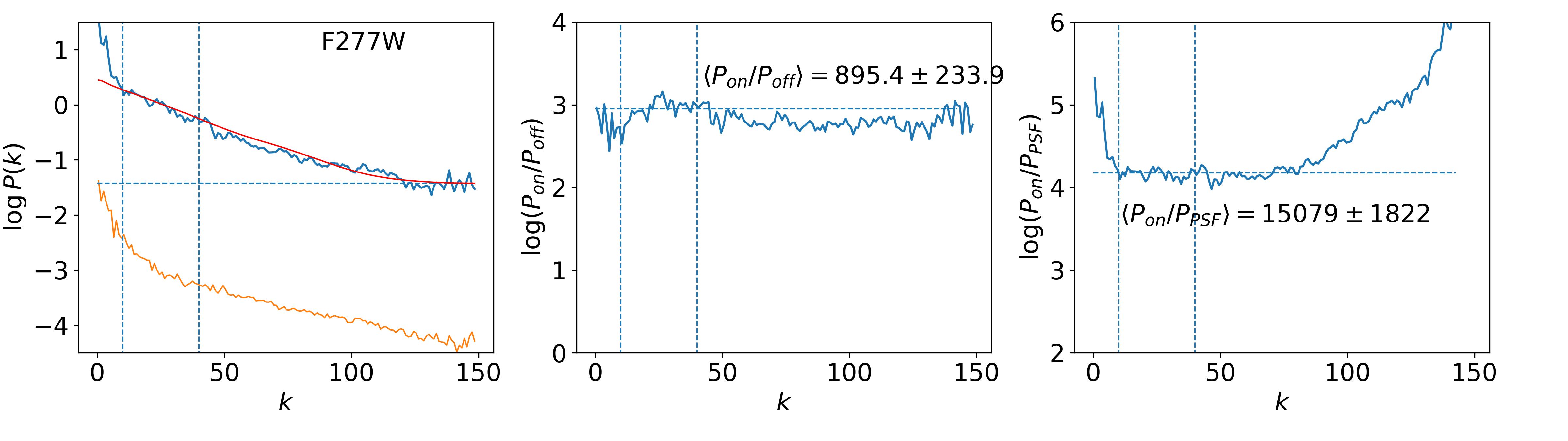}\vspace{-10pt}
    \caption{\small Power spectrum analyses of JWST data on NGC\,7317, an elliptical galaxy at ${\sim\,}$100\,Mpc, in the F150W (top row) and F277W (bottom row) passbands. Left panels: power spectra of the central masked region after galaxy subtraction (blue solid curve), the scaled PSF (red curve), and an empty region away from the galaxy (orange curve); dotted blue lines show the range of the starting waveneumbers $k$ for the fits. 
    Middle panels: ratio of the power spectra on- and off-galaxy; the structure in this ratio for F150W may indicate a variable PSF in the combined image. Right: Uncorrected power spectrum normalization $P_0$ derived from the fits in the left panels as a function of the starting $k$ value (the fitted $P_0$ values here are insensitive to the starting~$k$).
 }   \label{fig:jwst_ps}
\end{figure*}

The characteristic SBF signal is clearly observed across all passbands, including mid-infrared bands that have not previously been used in SBF studies. The left panels of Figure~\ref{fig:jwst_ps} display the measured power spectra in the F150W band of the short-wavelength channel and the F277W band of the long-wavelength channel. The other panels show the ratio of the fluctuation amplitudes on- and off- the galaxy center and the $P_0$ coefficient of the PSF term in the power spectrum, using a star within the image as the reference PSF. Analysis of the SBF signal for this target is in progress, and further work is needed to improve the agreement with the PSF templates, but the signal is very strong, given the short exposure times for this distance.

With the forthcoming generation of wide-area imaging surveys such as Rubin/LSST, Euclid, and the Roman Community Surveys providing hundreds or even thousands of galaxies suitable for SBF measurements, it is essential that the SBF analysis should be carried out in a robust, streamlined, and automatic way. Historically, the limited number of available SBF measurement pipelines have been poorly automated and required extensive user intervention. This is especially true for the galaxy fitting and residual masking, as these steps must be optimized to ensure clean power spectra. However, it is clear that the method needs to be adapted to run in ``production mode" to deliver robust results with minimal human intervention. Multiple parallel efforts are ongoing in this regard.

The prospects for SBF measurements with AO-assisted observations on
30-40m telescopes are also exciting. Table \ref{tab:sbf_h0} reports the expectations for E-ELT/MICADO with the MORFEO AO module. With their extremely large apertures and diffraction-limited resolution $<0\farcs02$, such facilities should be able to measure SBF distances out to redshift $z\sim0.1$. However,  measuring SBF with AO-assisted instruments may be challenging because of the spatially and temporally varying PSF, significant overheads, and the requirement for accurate $k$-corrections at these distances \cite{jensen21}. Nevertheless, once the technique is optimized for 30m-class telescopes, a carefully selected set of targets could have a significant impact, potentially providing an independent direct detection of cosmic acceleration.

\section{Conclusions}

Over the last 30 years, the SBF method has become  one of the most powerful methods for measuring galaxy distances from the Local Group to the Hubble flow and constraining the value of $H_0$. Unlike other precision distance indicators (Cepheids, TRGB, SN Ia, masers), SBF measurements require only a modest investment of telescope time, especially when using high-resolution space-based data. SBF distances to large numbers of galaxies observed with HST using WFC3/IR have already constrained $H_0$ to 3.4\%, where the main source of uncertainty is the zero-point calibration. Now, with JWST/NIRCam, it becomes possible to directly calibrate SBF using TRGB distances to the same galaxies, opening a path to reduce the uncertainty on $H_0$ from SBF to below 2\%. The large aperture, superior resolution, PSF stability, and near-infrared capability of JWST also make it possible to measure robust SBF distances out to  $\sim\,$300~Mpc. This will reduce the uncertainty from possible large-scale bulk motions and vastly increase the number of optimal targets in range of the method. Other forthcoming facilities such as Euclid, Roman, Rubin, and the ELTs also hold great potential for the method. With an improved TRGB-based calibration and samples of hundreds, or even thousands, of SBF distances reaching out to hundreds of Mpc, a fully independent measurement of $H_0$ to a precision approaching 1\% will be in reach by the end of he current decade.

\begin{table*}
\caption{Maximum Distances and Numbers of Candidates for SBF Analysis}
\centering
\begin{tabular}{lccccccc}
\hline \hline
 Telescope & Band & ~~Depth$^{(a)}$  & GCLF &  GCLF  &  $d_\mathrm{Max}$ & Optimal/good  & $\delta H_0/H_0$  (\%)   \\
 (constraint)          &      &  (mag)           & peak$^{(b)}$  &  offset$^{(c)}$   &  (Mpc)  & candidates   &  optimal/good \\
 (1)  & (2)  & (3) &  (4)  &  (5) &  (6) &  (7) &  (8)   \\
\hline 
Rubin (Dec$\leq15^{\circ}$)    & $i$          & 26.8 & $-$8.5 & $+$0.5 &  70  &  50/1500   &  1.2/$<$0.5  \\  
 Euclid ($|b|\geq23^{\circ}$)   & $H_{Euclid}$ & 24.0 & $-$8.6 & $-$1.5 &  55  &   20/600   & 1.0/$<$0.5  \\  
Roman$^{(e)}$     & $J_{F129}$    & 28.0 & $-$8.3 & $-$1.0 &  230 &  1500/$>$5000& 0.7/0.5$^{(g)}$  \\  
Roman$^{(e)}$     & $H_{F158}$   & 28.0 & $-$8.6 & $-$1.5 &  300 &  1600/$>$5000& 0.7/0.5$^{(g)}$   \\ 
JWST$^{(e)}$      & $J_{F115W}$  & 27.5 & $-$8.3 & $-$1.0 &  180 &  900/$>$5000 & 0.9/0.5$^{(g)}$  \\ 
JWST$^{(e)}$      & $H_{F150W}$  & 27.7 & $-$8.6 & $-$1.5 &  290 &  1600/$>$5000 & 0.7/0.5$^{(g)}$  \\  
JWST$^{(e)}$      & $K_{F200W}$  & 27.8 & $-$8.1 & $-$2.0 &  310 &  1600/$>$5000 & 0.7/0.5$^{(g)}$  \\ 
ELT$^{(f)}$ (Dec$\leq 0^{\circ}$)     & $K$          & 28.5 & $-$8.1 & $-$2.0 &  400 &  750/$>$4000 & 1.0/0.7$^{(g)}$  \\  
\hline  \hline
\multicolumn{8}{l}{
  \begin{minipage}{11.7cm}%
  \vspace{6pt}
\scriptsize $(a)$ Nominal $5\sigma$ point-source depth accounting for host galaxy background.\\[3pt]
\scriptsize $(b)$ Peak, or turnover, magnitude (AB) of the GCLF in each band. \\[3pt]
\scriptsize $(c)$ Offset of the required detection limit with respect to the GCLF peak (see text).\\[3pt]
\scriptsize $(d)$ Estimated number of optimal/suitable SBF targets based on 2MRS \cite{huchra12}.\\[3pt]
\scriptsize $(e)$ $5\sigma$ depth in one hour integration time.\\[3pt]
\scriptsize $(f)$ Depth derived from the preliminary ELT/MICADO exposure time calculator, assuming a $S/N=5$ in one hour in multi-conjugate adaptive optics mode.\\[3pt]
\scriptsize $(g)$ Assuming that a fraction of $\sim2\%$ of possible candidates are indeed observed.\\[3pt]
    \end{minipage}
    }
\end{tabular}\label{tab:sbf_h0}
\end{table*}

\newpage
{\footnotesize
\bibliography{mik_new}{}
\bibliographystyle{plain}
}

\end{document}